\documentclass[aps,prd,showpacs,preprintnumbers,superscriptaddress,nofootinbib,twocolumn]{revtex4}
\usepackage[dvips]{graphicx}
\usepackage{bm,latexsym,amsmath,amssymb,amsfonts,mathrsfs}
\usepackage{color}
\input{colordvi.tex}
\newcommand*{\D}{{\rm d}}
\newcommand*{\DD}{{\rm D}}
\newcommand*{\curl}{{\rm curl}\,}
\newcommand*{\cH}{{\cal H}}
\newcommand*{\mpl}{M_{\rm Pl}}
\begin{document}

\title{The Music of the Aetherwave - B-mode Polarization in Einstein-Aether Theory}

\author{Masahiro~Nakashima}
\email[Email: ]{nakashima"at"resceu.s.u-tokyo.ac.jp}
\affiliation{Department of Physics, Graduate School of Science, The University of Tokyo, Tokyo 113-0033, Japan}
\affiliation{Research Center for the Early Universe (RESCEU), Graduate School of Science, The University of Tokyo, Tokyo 113-0033, Japan}

\author{Tsutomu~Kobayashi}
\email[Email: ]{tsutomu"at"resceu.s.u-tokyo.ac.jp}
\affiliation{Research Center for the Early Universe (RESCEU), Graduate School of Science, The University of Tokyo, Tokyo 113-0033, Japan}

\begin{abstract}
We study how the dynamical vector degree of freedom in modified gravity affects the CMB B-mode polarization in terms of the Einstein-aether theory. In this theory, vector perturbations can be generated from inflation, which can grow on superhorizon scales in the subsequent epochs and thereby leaves imprints on
the CMB B-mode polarization.  
We derive the linear perturbation equations 
in a covariant formalism, and compute the CMB B-mode polarization using the CAMB code modified so as to incorporate the effect of the aether vector field.
We find that the amplitude of the B-mode signal from the aether field can surpass the contribution from the inflationary gravitational waves for a viable range of model parameters. We also give an analytic argument explaining the shape of the spectrum based on the tight coupling approximation. 
\end{abstract}

\preprint{RESCEU-4/11}
\maketitle

\section{Introduction}

It is widely believed that
the present Universe is dominated by the two dark components:
cold dark matter and dark energy.
Dark matter plays an essential role
in explaining galaxy rotation curves and
in structure formation, while dark energy is
presumably responsible for the current cosmic acceleration.
The presence of the dark components is thus
perceived through the gravitational interaction,
having not yet identified what they really are.
It is therefore legitimate to
think of these major mysteries of today's cosmology
as a mystery of gravity.
This motivates us to explore long-distance modification of the gravitational law,
asking to what extent general relativity (GR) is correct on cosmological scales.

Modification of gravity is most commonly made by adding
an extra scalar degree of freedom \'{a} la Brans-Dicke gravity~\cite{bd}.
In recent years, various refined models of scalar-tensor gravity
have been proposed which would be an alternative to material dark energy
while being consistent with solar-system constraints.
They include chameleon $f(R)$~\cite{cham, fr} and Galileon~\cite{gal, gde} theories,
and have been tested against cosmological observations~\cite{tests}.
It is also possible to modify the spin-2 sector
as in massive gravity~\cite{massiveg} and bi-gravity theories~\cite{bigra}.

In this paper, we are going to consider
a hypothetical {\em vector} degree of freedom of gravity.
Specifically, we shall focus on the Einstein-aether (EA) theory
proposed by Jacobson and Mattingly~\cite{Jacobson:2000xp},
in which a fixed norm vector field
with a Lorentz-violating vacuum expectation value
takes part in the gravitational interaction.
The effect of the aether on the cosmological background was clarified in~\cite{Carroll:2004ai}.
The scalar cosmological perturbations
in the EA theory and their impact on the cosmic microwave background (CMB)
temperature anisotropy have been studied in~\cite{Lim:2004js,Li:2007xw,Zuntz:2008zz,Zuntz:2010jp}. 
Recently, Armendariz-Picon {\it et al}.~\cite{Garriga} performed a comprehensive analysis on cosmological perturbations in the EA theory.
See also Refs.~\cite{Eling:2006df,Eling:2007xh,Eling:2006ec,Tamaki:2007kz}
for the other aspects of the EA theory, such as spherically symmetric solutions
and compact objects.
Interestingly, it was recently pointed out that
the healthy extension~\cite{blas}
of Horava's quantum theory of gravity~\cite{horava}
reduces to a special case of the EA theory at low energies~\cite{Jacobson:2010mx}.
Cosmological perturbations in the healthy extension of Horava gravity
were studied in~\cite{cphealthy}.


The purpose of the present paper is to clarify the impact of
the aether vector field on the CMB polarization.
The CMB polarization arises from all the three types of cosmological perturbations, {\it i.e.},
scalar, vector, and tensor perturbations.
Among them, as pointed out by~\cite{Hu:1997hp}, the vector perturbations most effectively generate the B-mode polarization. 
However, the effect of vector perturbations has been less investigated
because the vector mode decays unless sourced, {\em e.g.}, by topological defects~\cite{Pen:1997ae,Seljak:2006hi},
while the scalar and tensor modes are certainly generated from inflation~\cite{Starobinsky:1979ty}.
Other possible ways of seeding vector perturbations
include the neutrino anisotropic stress~\cite{Lewis-Vector},
the second-order effect~\cite{2ndvec1,2ndvec2,2ndvec3,2ndvec4,Pitrou:2008hy},
and primordial magnetic fields generated somehow~\cite{Lewis:2004ef}.
Modifying the vector sector of gravity offers
a yet another possibility of producing vector perturbations,
leaving a unique signature in the CMB polarization due to nontrivial dynamics of the aether field.

The paper is organized as follows.
In the next section we introduce the EA theory and the basic equations. In Sec.~III,
we describe the dynamics of
vector perturbation in the EA theory
using the covariant approach, in order to incorporate the aether vector field
into the CAMB code.
We then specify the initial conditions for the perturbation evolution in Sec.~IV.
Our numerical results are presented in Sec.~V.
In Sec.~VI, we examine the spectrum shape in an analytic approach using the tight coupling approximation.
We draw our conclusions in Sec.~VII.
We will use the sign convention $(+ - - - )$.

\section{Einstein-aether theory}

The action of the EA theory is given by~\cite{Jacobson:2000xp}
\begin{eqnarray}
{\cal S}=\frac{\mpl^2}{2}\int\D^4x\sqrt{-g}[{\cal R}+{\cal L}_A]+{\cal S}_m,
\end{eqnarray}
where
\begin{eqnarray}
{\cal L}_A&=&-[c_1\nabla_aA^b\nabla^aA_b+c_2(\nabla_bA^b)^2+c_3\nabla_aA^b\nabla_bA^a
\nonumber\\
&&+c_4A^aA^b\nabla_aA^c\nabla_bA_c]+\lambda(A_bA^b-1)
\end{eqnarray}
is the Lagrangian of the aether field and ${\cal S}_m$ is the action of ordinary matter.
It is assumed that the aether is not coupled to the matter field directly.

Variation with respect to the metric leads to the Einstein equations
\begin{eqnarray}
{\cal R}_{ab}-\frac{1}{2}g_{ab}{\cal R}=T_{ab}+\kappa \tau_{ab},
\end{eqnarray}
where ${\cal R}_{ab}$ is the Ricci tensor, $\kappa=\mpl^{-2}$,
\begin{eqnarray}
T_{ab}:=\frac{1}{2}{\cal L}_Ag_{ab}-\frac{\delta{\cal L}_A}{\delta g^{ab}}
\end{eqnarray}
is the energy-momentum tensor of the aether,
and $\tau_{ab}$ is the energy momentum tensor of ordinary matter.
Explicitly, we have
\begin{eqnarray}
T_{ab}&=&
c_1\left[
(\nabla_aA^c)(\nabla_bA_c)
-(\nabla_cA_a)(\nabla^cA_b)
\right]
\nonumber\\&&
+c_1\nabla_c\left[
A^c\nabla_{(a}A_{b)}+(\nabla^cA_{(a})A_{b)}
-A_{(a}\nabla_{b)}A^c
\right]
\nonumber\\&&
+c_2g_{ab}\nabla_c\left(A^c\nabla_dA^d\right)
\nonumber\\&&
+c_3\nabla_c\left[
A^c\nabla_{(a}A_{b)}-(\nabla^cA_{(a})A_{b)}
+A_{(a}\nabla_{b)}A^c
\right]
\nonumber\\&&
-c_4A^cA^d(\nabla_cA_a)(\nabla_dA_b)
\nonumber\\&&
-c_4\nabla_c\left[
A_aA_bA^d\nabla_dA^c-2A^cA^d(\nabla_dA_{(a})A_{b)}
\right]
\nonumber\\&&
+\frac{1}{2}{\cal L}_A g_{ab}
+\lambda A_aA_b.
\end{eqnarray}
Ordinary matter includes photons, baryons, etc., so that we write
$\tau_{ab}=\sum_i \tau_{ab}^{(i)}$, where $i$ labels different components.
Note that ${\cal L}_A$ is taken to be a functional of $A^a$ rather than $A_a$.
Variation with respect to $A^a$ yields the equation of motion for the aether:
\begin{eqnarray}
&&c_1 \Box A_a+c_2\nabla_a\nabla_bA^b+c_3\nabla_b\nabla_aA^b
\nonumber\\&&
+c_4\left[\nabla_b\left(A^bA^c\nabla_cA_a\right)
-A^b(\nabla_bA_c)\nabla_aA^c
\right]=-\lambda A_a.
\nonumber\\ \label{AEEOM}
\end{eqnarray}
Finally, variation with respect to the Lagrange multiplier $\lambda$
gives the fixed norm constraint
\begin{eqnarray}
A_aA^a=1.\label{fixtnorm}
\end{eqnarray}
It is convenient to use the following abbreviations:
\begin{eqnarray}
&& c_{13} = c_1+c_3,
\quad
c_{14}=c_1+c_4,
\quad  \nonumber \\
&& \alpha = c_1+3c_2+c_3,
\quad 
c_{123} = c_{1}+c_{2}+c_{3}
\end{eqnarray}

\section{Covariant approach}

To describe background cosmology and the evolution of vector perturbations,
we employ the covariant equations obtained by
the method of $3+1$ decomposition.
We begin with splitting physical quantities with respect to observer's 4-velocity $u^a$.
Following the usual procedure, the projection tensor is defined as
$h_{ab}:=g_{ab}-u_au_b$ and the covariant spatial derivative $\DD_a$ acting on
a tensor field $T^{b\cdots }_{c\cdots }$ is defined as
$\DD^aT^{b\cdots }_{c\cdots } := h^a_ih^b_j\cdots h^k_c\cdots \nabla^i T^{j\cdots }_{k\cdots }$.
The energy-momentum tensors of ordinary matter and the aether field
are decomposed respectively as
\begin{eqnarray}
\tau_{ab}^{(i)}&=&\rho^{(i)} u_a u_b-p^{(i)} h_{ab}+2q^{(i)}_{(a}u_{b)}+\pi^{(i)}_{ab},
\\
T_{ab}&=&
\tilde \rho u_a u_b-\tilde p h_{ab}+2\tilde q_{(a}u_{b)}+\tilde\pi_{ab},
\end{eqnarray}
while $\nabla_au_b$ is decomposed as
\begin{eqnarray}
\nabla_au_b=\frac{1}{3}\theta h_{ab}+\sigma_{ab}+\omega_{ab}-u_a\dot u_b.
\end{eqnarray}
Here, $\sigma_{ab}:=\DD_{(a}u_{b)}-(1/3)\nabla_cu^c h_{ab}$ is the shear tensor,
$\omega_{ab}:=\DD_{[a}u_{b]}$ is the vorticity, $\theta:=\nabla_au^a$ is the expansion,
and the overdot denotes time derivative $\dot{}:=u^a\nabla_a$.
The expansion may be written as $\theta=3\dot S/S$, where $S$ is the averaged scale factor.

The conservation equations for the matter energy-momentum tensor imply
\begin{eqnarray}
&&\dot\rho +\theta\left(\rho +p \right)+\DD^a q_a =0,
\\
&&\dot q_a+\frac{4}{3}\theta q_a+\left(\rho+p \right)
\dot u_a-\DD_a p+\DD^b\pi_{ab}=0,
\end{eqnarray}
where $\rho=\sum_i\rho^{(i)}$, $p=\sum_ip^{(i)}$,
$q_a=\sum_iq_a^{(i)}$, and $\pi_{ab}=\sum_i\pi^{(i)}_{ab}$.
We also have the corresponding conservation equations for the aether:
\begin{eqnarray}
&&\dot{\tilde\rho}+\theta\left(\tilde\rho+\tilde p\right)+\DD^a\tilde q_a =0,
\\
&&\dot{\tilde{q}}_a+\frac{4}{3}\theta \tilde{q}_a+(\tilde\rho+\tilde p) \dot u_a-\DD_a \tilde p
+\DD^b\tilde \pi_{ab}=0.
\label{mom-conserv-AE}
\end{eqnarray}

Let us first consider the homogeneous and isotropic limit.
In this limit, the constraint~(\ref{fixtnorm}) implies $A^a=u^a$.
Substituting this to the equation of motion~(\ref{AEEOM}), one finds
\begin{eqnarray}
\lambda = \frac{c_{13}}{3}\theta^2-c_2\dot\theta.\label{Lm:fixed}
\end{eqnarray}
The energy density and pressure of the aether are then given by
\begin{eqnarray}
\tilde\rho &=& c_2\left(\dot\theta+\theta^2\right)+\lambda
-\frac{\alpha}{6}\theta^2
\nonumber\\&=&
\frac{\alpha}{6}\theta^2,
\\
\tilde p&=&-\frac{\alpha}{6}\left(2\dot\theta+\theta^2\right).
\end{eqnarray}
It can be seen that $\tilde\rho$ and $\tilde p$ are expressed in terms of the expansion $\theta$,
and they are the same as what appear in the left hand side of the Einstein equations.
This means that the effect of the aether on the background evolution is
just to renormalize the gravitational constant:
$\kappa \to\tilde\kappa:=(1-\alpha/2)^{-1}\kappa$~\cite{Carroll:2004ai}.
The background equations are thus given by
\begin{eqnarray}
{\cal H}^2 &=& \frac{\tilde\kappa}{3}S^2\rho, \label{bg-fried}
\\
{\cal H}^{\prime} &=& -\frac{\tilde\kappa}{6}S^{2}(\rho+3p),
\end{eqnarray}
where we have introduced the comoving Hubble parameter, ${\cal H}:=S\theta/3$,
and the derivative with respect to the conformal time, $':=S u^a\nabla_a$.

In order for the background Friedmann equation (\ref{bg-fried}) to have a solution, the condition
\begin{equation}
\alpha < 2 \label{background_stability}
\end{equation}
must be satisfied for positive matter energy density and positive $\kappa$. On the other hand,
the effective gravitational constant on small scales $\kappa_{N}$ is also different from the bare one: $\kappa_{N} = (1+c_{14}/2)^{-1}\kappa$~\cite{Carroll:2004ai}.
The difference between these two effective gravitational constants is constrained by nucleosynthesis as $|1-\tilde\kappa/\kappa_{N}| < 10 \%$. In terms of the aether parameters, this constraint is roughly expressed as 
\begin{equation}
c_{14}+\alpha \lesssim 0.2.  
\end{equation}  
Note that this is trivially satisfied if we consider the special case $\alpha=-c_{14}$. Actually, this is the special combination for evading
the existing observational constraints on $c_{i}$
(see Appendix A), and we will often use this case later. 

Having studied the background effect of the aether,
we then move on to the dynamics of vector perturbations.
As employed in~\cite{Lewis-Vector},
we choose $u_a$ to be hypersurface orthogonal,
so that $\curl u_b=0\;\Rightarrow\; \dot u_b=0$ at linear order.
This simplifies the following analysis.

At linear order, the aether field can be written as
\begin{eqnarray}
A_b = u_b+\DD_b V^{(s)} + V_b,
\end{eqnarray}
where $V^{(s)}$ and $V_b$ are first order quantities.
$V^{(s)}$ corresponds to a scalar perturbation which we do not consider in this paper,
while $V_b$ a vector perturbation that satisfies $\DD_bV^b=0$.
The fixed norm constraint~(\ref{fixtnorm}) leads to $u_bV^b=0$, i.e., $V^b$ is a spatial vector.
Since
$
\nabla_aV_b=\DD_aV_b-(1/3)\theta V_a u_b+u_a \dot V_b
$,
we have
\begin{eqnarray}
\nabla_aA_b=\frac{1}{3}\theta h_{ab}
+\sigma_{ab}+\DD_aV_b-\frac{1}{3}\theta V_a u_b+u_a \dot V_b.
\end{eqnarray}
Substituting this to Eq.~(\ref{AEEOM}), we find, up to first order,
\begin{eqnarray}
&&
c_{13}\left[
\frac{\theta^2}{3}u_a-\DD^b\sigma_{ab}
+\frac{1}{3}\left(\dot\theta+\theta^2\right)V_a
\right]
-c_1\DD^2V_a
\nonumber\\&&\;\;
-c_3\DD_b\DD_aV^b
-c_2\dot\theta u_a-c_{14}\left(\dot \chi_a+\frac{2}{3}\theta\chi_a\right)
\nonumber\\&&
=\lambda u_a+\lambda V_a,\label{AE:b+p}
\end{eqnarray}
where
\begin{eqnarray}
\chi_a=\dot V_a+\frac{1}{3}\theta V_a.
\end{eqnarray}
Since we are interested in vector perturbations,
we have dropped in the above the scalar perturbation $\DD_a\theta$.
Multiplying $u^a$ 
gives rise to the background equation which we have already derived.
Multiplying $h_b^{\;a}$, we obtain
\begin{eqnarray}
&&c_{14}\left(\dot\chi_a+\frac{2}{3}\theta\chi_a\right)=
\nonumber\\&&
-\left(c_{13}\DD^b\sigma_{ab}+c_1\DD^2V_a+c_3\DD_b\DD_aV^b\right)
+\frac{\alpha}{3}\dot\theta V_a,
\end{eqnarray}
where we used Eq.~(\ref{Lm:fixed}).
At linear order scalar-type quantity $\tilde\rho$ and $\tilde p$
have the same expression as in the background.
The heat-flux vector $\tilde q_a$ and the anisotropic stress $\tilde\pi_{ab}$ of the aether
are given respectively by
\begin{eqnarray}
\tilde q_a&=&c_{14}\left(\dot\chi_a+\frac{2}{3}\theta\chi_a\right)
+\frac{c_1-c_3}{2}\left(
\DD^2 V_a-\DD_b\DD_aV^b\right)
\nonumber\\&&
-\frac{c_{13}}{3}\left(\dot\theta+\theta^2\right)V_a
+\lambda V_a,
\nonumber\\
&=&-c_{13}\DD^b\left[
\sigma_{ab}+\DD_{(a}V_{b)}\right],
\label{til-q}
\\
\tilde\pi_{ab}&=&c_{13}\left\{
\dot\sigma_{ab}+\theta\sigma_{ab}
+\left[\DD_{(a}V_{b)}\right]\dot{}+\theta \DD_{(a}V_{b)}
\right\},
\end{eqnarray}
where we used Eq.~(\ref{Lm:fixed}) to remove $\lambda$.
One can check that the momentum conservation equation~(\ref{mom-conserv-AE})
is automatically satisfied.

Combining Eq.~(\ref{til-q}) with
the momentum constraint equation,
\begin{eqnarray}
\DD^b\sigma_{ab}=\kappa q_a+\tilde q_a,
\end{eqnarray}
we obtain
\begin{eqnarray}
\DD^b\sigma_{ab}=\frac{1}{1+c_{13}}\left[\kappa q_a-c_{13}\DD^b\DD_{(a}V_{b)}\right].
\end{eqnarray}
Note in passing that
$\DD^b\DD_aV_b=0$ at linear order.

To proceed further, it is convenient to introduce
the transverse eigenfunctions.
The definitions and the basic properties of the eigenfunctions
are presented in~\cite{Lewis-Vector}.
In terms of the eigenfunctions the vector perturbations can be expanded as
\begin{eqnarray}
V_a =\sum VQ_a^{\pm},
\quad
\sigma_{ab}=\sum \frac{k}{S}\sigma Q_{ab}^{\pm},
\nonumber\\
q_a^{(i)} = \sum q_{i} Q_a^{\pm},\quad
\pi_{ab}^{(i)}=\sum \Pi_{i} Q_{ab}^{\pm},
\end{eqnarray}
where $k$ is the eigenvalue.
In terms of the harmonic coefficients,
our perturbation equations are written as
\begin{eqnarray}
&&q'+4{\cal H}q  +\frac{k}{2}\Pi=0,
\\
&&k^2\sigma=\frac{1}{1+c_{13}}\left(
2\kappa S^2 q-c_{13}k^2V
\right),\label{per-const}
\\
&&
c_{14}\left[V''+2{\cal H}V'+\left({\cal H}^2+{\cal H}'\right)V\right]
\nonumber\\&&\qquad\quad
+\alpha\left({\cal H}^2-{\cal H}'\right)V
+c_1k^2 V
=-\frac{c_{13}}{2}k^2\sigma,\label{per-ae}
\end{eqnarray}
where $q:=\sum_iq_i$ and $\Pi:=\sum_i\Pi_i$.

Equation~(\ref{per-ae}) shows that the fluctuation of the aether
obeys the wave equation which is similar to the
evolution equation for cosmological tensor perturbations.
The crucial difference is the effective mass term which is dependent on
the expansion rate ${\cal H}$ and the model parameters.
The fluctuation of the aether is related to $\sigma$ via the momentum constraint,
which in turn translates to the magnetic Weyl tensor
$H_{ab}$ via $H_{ab} = {\rm curl}\;\sigma_{ab}$, and thus produces
perturbations of geometry.
If we neglect the matter contents, the above equations reduce to those derived in~\cite{Garriga}.

The velocity $v_{i}$ of each fluid component is given by
$v_{i}=q_{i}/(\rho^{(i)}+p^{(i)})$.
The baryons are coupled to photons via Thomson scattering.
The baryon velocity $v_b$ obeys
\begin{eqnarray}
v_b'+{\cal H}v_b=-\frac{\rho_\gamma}{\rho_b}Sn_e\sigma_T\left(\frac{4}{3}v_b-I_1\right),
\end{eqnarray}
with $I_1=4v_\gamma /3$.

We now replicate the photon multipole equations
and the polarization multipole equations for vectors presented in~\cite{Lewis-Vector,Challinor:1998xk,Challinor:2000as}.
The photon multipole equations for vectors are
\begin{widetext}
\begin{eqnarray}
I_\ell'+k\frac{\ell}{2\ell+1}\left(\frac{\ell+2}{\ell+1}I_{\ell+1}-I_{\ell-1}\right)
=-S n_e\sigma_T\left(
I_\ell-\frac{4}{3}\delta_{\ell 1} v_b-\frac{2}{15}\zeta\delta_{\ell 2}
\right)+\frac{8}{15}k\sigma\delta_{\ell 2}, \label{eq:photon-multipole}
\label{mp-I}
\end{eqnarray}
where $\zeta:=3I_2/4-9E_2/2$ and $I_2=\Pi_\gamma/\rho_\gamma$,
while the polarization multipole equations for vectors are
\begin{eqnarray}
&&E_{\ell}^{\pm\prime} +\frac{(\ell+3)(\ell+2)\ell(\ell-1)}{(\ell+1)^{3}(2\ell+1)}kE_{\ell+1}^\pm-\frac{\ell}{2\ell+1}kE_{\ell-1}^\pm
-\frac{2}{\ell(\ell+1)}kB_{\ell}^{\pm} =-Sn_{e}\sigma_{T}\left(
E_\ell^\pm-\frac{2}{15}\zeta^\pm\delta_{\ell 2}\right), \label{eq:El-multipole}
\\
&&B_\ell^{\pm\prime}+\frac{(\ell+3)(\ell+2)\ell(\ell-1)}{(\ell+1)^3(2\ell+1)}kB_{\ell+1}^\pm-\frac{\ell}{2\ell+1}kB_{\ell-1}^\pm
+\frac{2}{\ell(\ell+1)}kE_\ell^\pm =-Sn_e\sigma_T
B_\ell^\pm, \label{eq:Bl-multipole}
\end{eqnarray}
\end{widetext}
where $E_\ell$ and $B_\ell$ are moments of the E and B
polarization.\footnote{Here we have corrected the typo found in~\cite{Lewis-Vector}.}
The integral solutions to these equations are given in~\cite{Lewis-Vector}.
We will only use the solution for $B_{\ell}$ in the following discussion:
\begin{eqnarray}
B_{\ell}(\eta_0)=-\frac{\ell-1}{\ell+1}\int^{\eta_0}\D\eta \,\dot{\tau} e^{-\tau}\Psi_\ell[k(\eta_0-\eta)]\zeta, \label{eq:Bl-sol}
\end{eqnarray}
where $\Psi_\ell(x):=\ell j_\ell(x)/x$ corresponding to the projection function $\beta_{\ell}^{(1)}$ in the total angular momentum approach \cite{Hu:1997hp}, and $\tau$ represents the optical depth: $\tau := \int^{\eta_{0}}\D\eta Sn_{e}\sigma_{T}$.
The neutrino multipole equations are of the form~(\ref{mp-I}) without
the Thomson scattering terms.

\section{Initial conditions}

We solve the relevant set of equations at early times
in order to clarify the initial conditions.
This is done by a series expansion in terms of the conformal time $\eta$,
following~\cite{Lewis-Vector} but now taking into account the presence of
the aether.

First, by invoking the tight coupling approximation for the baryons and photons
it is easy to obtain
\begin{eqnarray}
v_\gamma\simeq v_b\simeq \frac{v_0}{1+R},
\end{eqnarray}
where $v_0$ is the initial value and $R=3\rho_b/4\rho_\gamma$.

The Friedmann equation gives the scale factor in terms of the conformal time as
\begin{eqnarray}
S=\frac{\Omega_R}{\Omega_m}\left(\omega\eta+\frac{1}{4}\omega^2\eta^2+\cdots\right),
\end{eqnarray}
where $\Omega_R:=\Omega_\gamma+\Omega_\nu$ and $\omega:=\Omega_m{\cal H}_0/\sqrt{\Omega_R}$. The definitions of $\Omega_i$ are the standard ones, namely, densities in units of the critical density.

Neglecting the ${\cal O}(k^2)$ terms at early times in the radiation-dominated era,
the perturbed equation of motion for the aether~(\ref{per-ae}) can be solved to give
\begin{eqnarray}
V={\cal A}_k\eta^\nu\left[1-\left(1-\frac{\nu}{2}\right)\frac{\omega\eta}{4}\right]
+{\cal A}_k^{(d)} \eta^{-1-\nu}+\cdots, \label{sh-sol-V}
\end{eqnarray}
where ${\cal A}_k$ and ${\cal A}_k^{(d)}$ may depend on $k$, and
\begin{eqnarray}
\nu :=\frac{-1+\sqrt{1-8\alpha/c_{14}}}{2}.
\end{eqnarray}
As we are interested in a non-decaying regular mode,
we assume that $\alpha/c_{14}\le 0$ and set ${\cal A}^{(d)}_k=0$.
In order for scalar isocurvature modes not to grow, the condition
$\alpha/ c_{14}\ge -1$ must be imposed~\cite{Garriga}.
We advocate this constraint and consider the range
\begin{eqnarray}
0\le \nu\le 1.
\end{eqnarray}

The neutrino multipole equations read
\begin{eqnarray}
I_1^{(\nu)\prime}+\frac{k}{2}I_2^{(\nu)}=0,\quad
I_2^{(\nu)\prime}-\frac{2k}{5}I_1^{(\nu)}=\frac{8}{15}k\sigma,\label{nu-multi}
\end{eqnarray}
where $I_1^{(\nu)}=4v_\nu/3$ and $I_2^{(\nu)}=\Pi_\nu/\rho_\nu$.
We use the multipole equations~(\ref{nu-multi}) and
the momentum constraint~(\ref{per-const})
to get the following early time solution:
\begin{eqnarray}
\sigma &=& {\cal B}_k\left(1-\frac{15}{2}\frac{\omega\eta}{4R_\nu^*+15}\right)
\nonumber\\&&\;\;
-\frac{\nu^*}{\nu^*+4R_\nu^*}\frac{c_{13}}{1+c_{13}}{\cal A}_k\eta^\nu,
\\
v_\gamma&=&\frac{{\cal B}_k}{4}\frac{4R_\nu^*+5}{R_\gamma^*}\left(1-
\frac{3R_b}{4R_\gamma }\omega\eta \right),
\\
v_\nu&=&-\frac{{\cal B}_k}{4}\frac{4R_\nu^*+5}{R_\nu^*}+{\cal O}(\eta^2),
\\
\frac{\Pi_\nu}{\rho_\nu}&=&-\frac{2{\cal B}_k}{3}\frac{k\eta}{R_\nu^*}
\left(1+\frac{3R_\nu^*}{15+4R_\nu^*}\omega\eta\right)
\nonumber\\&&\;\;
-\frac{8}{15(1+\nu)}\frac{\nu^*}{\nu^*+4R_\nu^*}\frac{c_{13}}{1+c_{13}}{\cal A}_kk\eta^{1+\nu}.
\end{eqnarray}
In the above we defined
\begin{eqnarray}
R^*_i:=\frac{1-\alpha/2}{1+c_{13}}R_i,
\quad \nu^*:=\frac{5}{2}(1+\nu)(2+\nu),
\end{eqnarray}
where $R_i$'s are defined in the same way as in~\cite{Lewis-Vector}:
$R_\nu:=\Omega_\nu/\Omega_R$, $R_\gamma=\Omega_\gamma/\Omega_R$,
and $R_b=\Omega_b/\Omega_m$.
The mode associated with ${\cal B}_k$
is identified as the regular vector
mode in the presence of the neutrino anisotropic stress~\cite{Lewis-Vector}.
Since ${\cal B}_k$ may be fixed independently of the effect of the aether ${\cal A}_k$, we discard this mode for clarity and focus on the
initial condition with ${\cal A}_k\neq 0$ and ${\cal B}_k=0$.

Once the inflation model and the subsequent reheating history are specified,
one can determine the primordial spectrum of the vector perturbation
and hence ${\cal A}_k$.
During inflation with $\epsilon:=1-{\cal H}'/{\cal H}^2 =$ const, one finds, on superhorizon scales,
that~\cite{Garriga}
\begin{eqnarray}
V\sim \frac{1}{M_{\rm Pl}}\frac{(-\eta)^{1/2}}{a}(-k\eta)^{(n_{v}-3)/2},
\end{eqnarray}
where
\begin{eqnarray}
n_{v}:=3-\sqrt{1-\frac{\alpha}{c_{14}}\frac{4\epsilon }{(1-\epsilon)^2}}.
\end{eqnarray}
For $-1\le \alpha/c_{14}\le 0$, we have $2-2\epsilon/(1-\epsilon)\le n_{v}\le 2$.
At the end of inflation, $\eta=\eta_{\rm e}$,
we have the estimate
\begin{eqnarray}
k^{3/2}V\sim \frac{H}{M_{\rm Pl}}\left(\frac{k}{k_{\rm e}}\right)^{n_{v}/2},
\label{priV}
\end{eqnarray}
where $H$ is the inflationary Hubble scale and $k_{\rm e}^{-1}$
corresponds to the horizon scale at $\eta=\eta_{\rm e}$.
The factor $k^{n_{v}/2}$ reflects the fact that
$V$ decays during inflation if $n>0$.
The amplitude may further change from (\ref{priV}) during the reheating stage,
and hence the primordial amplitude depends also on
the detailed history of reheating. 
In our actual calculation, we simply assume that
\begin{eqnarray}
{\cal A}_k={\cal A}_0 k^{(n_{v}-3)/2},
\end{eqnarray}
where ${\cal A}_0$ is a constant.

As was already derived in Eq.~(\ref{sh-sol-V}), $V$ grows on superhorizon scales,
$V\sim \eta^\nu$, in the radiation-dominated stage.
In the matter-dominated stage it turns out that
the superhorizon behavior is given by $V\sim \eta^{\nu_{\rm m}}$ with
\begin{eqnarray}
\nu_{\rm m}:=\frac{-3+\sqrt{1-24\alpha/c_{14}}}{2}.
\end{eqnarray}
Since $-1\le\nu_{\rm m}\le 1$, $V$ may grow or decay in the matter-dominated stage,
depending on $\alpha/c_{14}$. Therefore,
although the amplitude (\ref{priV}) (or ${\cal A}_0$) may be tiny as a consequence
of the decay of $V$ during inflation,
in the subsequent radiation- and matter-dominated stages $V$ can be amplified
on superhorizon scales,
leading to an observationally relevant aether perturbation.
The special case $\alpha=-c_{14}$ is an example of such situations, for which $\nu=\nu_{\rm m}=1$.

\section{Numerical Results}
\begin{figure}[t]
\centering
\includegraphics[keepaspectratio=true,width=9cm]{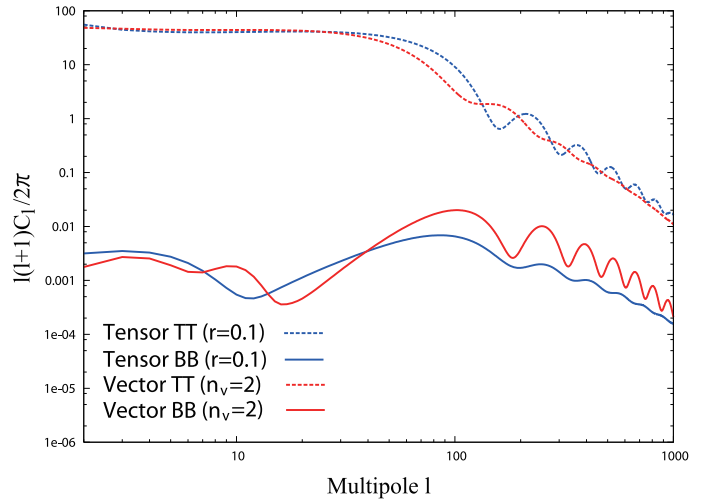}
\caption{
CMB B-mode polarization and temperature anisotropy power spectra in the EA theory. For comparison, those from the tensor perturbation in standard GR are also plotted in the case of the tensor-to-scalar ratio $r=0.1$. In this figure, $c_{1}=-0.2,c_{13}=-0.3,c_{14}=-\alpha=-0.2$, and dimensionless primordial power spectra are $\mathcal{P}_{V}\propto k^{n_{v}}$ and $\mathcal{P}_{T}\propto k^{0}$.
}
\label{fig:P48_ex1}
\end{figure}

Using all the ingredients derived
above and the CAMB code~\cite{Lewis:1999bs} modified so as to incorporate the presence of the aether, we have completed the numerical calculation for the B-mode polarization power spectra in the EA theory. An example of our numerical results is presented in the Fig.~\ref{fig:P48_ex1}.
For comparison, we show
contributions from inflationary gravitational waves in GR, assuming that
the tensor-to-scalar ratio is given by $r=0.1$.
In fact, the aether modifies the behavior of the tensor modes as well. 
However, the effects of the aether on the tensor modes are just to shift the location of the peak and to change the absolute amplitude of the primordial spectrum, and they are very small for small values of $c_{i}$. 
The amplitude ${\cal A}_0$ is adjusted so that the low-$\ell$ TT spectrum from the vector perturbation
has the same magnitude
as this primordial tensor contribution.
We see in this case that the BB spectrum in the EA theory
is larger than that from primordial tensor modes at $\ell\gtrsim 100$,
and hence the B-mode is potentially detectable in future CMB observations
aiming to detect $r={\cal O}(0.1)-{\cal O}(0.01)$~\cite{:2006uk}.
In plotting Fig.~\ref{fig:P48_ex1}, we chose $\alpha=-c_{14}$. In this case, it has been discussed in \cite{Garriga} that the TT power spectrum has roughly the same shape as the one from the inflationary gravitational waves. As one can see, Fig.~\ref{fig:P48_ex1} shows the same scalings for the two TT spectra.

\begin{figure}[t]
\centering
\includegraphics[keepaspectratio=true,width=9cm]{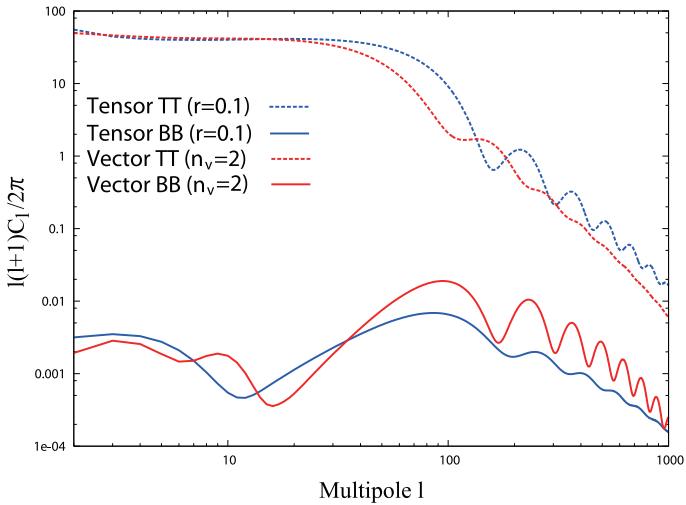}
\caption{
CMB B-mode polarization and temperature anisotropy power spectra in the EA theory. For comparison, those from the tensor perturbation in standard GR are also plotted in the case of the tensor-to-scalar ratio $r=0.1$. In this figure, $c_{1}=-0.019,c_{13}=-0.03,c_{14}=-\alpha=-0.0128$, and dimensionless primordial power spectra are $\mathcal{P}_{V}\propto k^{n_{v}}$ and $\mathcal{P}_{T}\propto k^{0}$.
}
\label{fig:good_parameterset}
\end{figure}

Using a more realistic parameter set evading all the existing constraints (see Appendix~A),
we plot the B-mode spectrum in Fig.~\ref{fig:good_parameterset}.
From this we conclude that, for a viable range of the model parameters,
the B-mode from the vector perturbation is potentially detectable in future CMB probes
even if its amplitude at the end of inflation is very small.
Note that the amplification of $V$ after inflation
is determined basically by the ratio between $\alpha$ and $c_{14}$.
This means that, even
if the model parameters $c_i$ are too small to discriminate the EA theory from GR
with the other observational and experimental tests,
the CMB B-mode polarization could be a powerful probe for the aether field.


\begin{figure}[t]
\centering
\includegraphics[keepaspectratio=true,width=8cm]{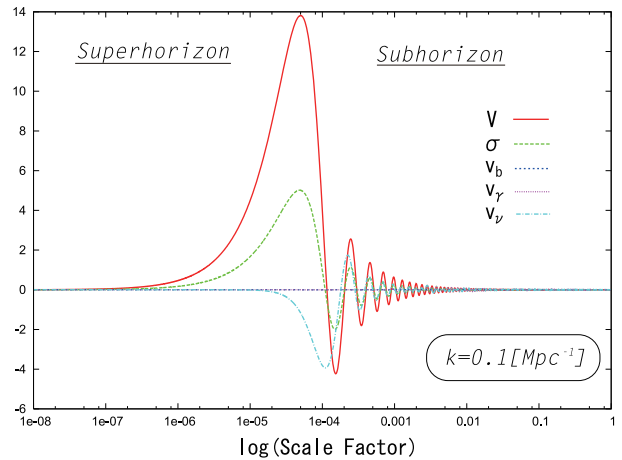}
\caption{
Evolution of each variable in a normal plot. The horizontal axis is the scale factor in log plot and $a_{0}=1$. In this figure, $c_{1}=-0.2,c_{13}=-0.3,c_{14}=-\alpha=-0.2$.
}
\label{fig:params1}
\end{figure}

\begin{figure}[t]
\centering
\includegraphics[keepaspectratio=true,width=8cm]{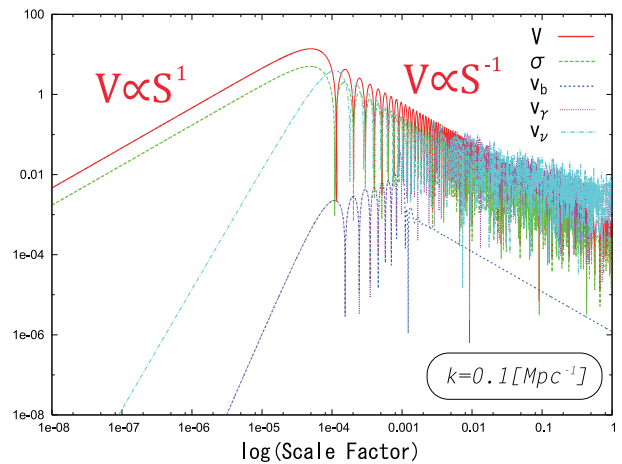}
\caption{
Evolution of each variable in a log plot. The horizontal axis is the scale factor in log plot and $a_{0}=1$. In this figure, $c_{1}=-0.2,c_{13}=-0.3,c_{14}=-\alpha=-0.2$.
}
\label{fig:params2}
\end{figure}

We show the evolution of each variable in a normal plot (Fig.~\ref{fig:params1}) and in a log plot (Fig.~\ref{fig:params2}). From these figures, we find that compared with the aether perturbation $V$ and the shear $\sigma$, matter components are negligibly small especially at early times.
It can be seen from Fig.~\ref{fig:params2} that
the growth rate of $V$ on superhorizon scales is given by
$V\propto S$ for $\alpha=-c_{14}$. This confirms the early time solution
derived in the previous section.


\section{Analytic Estimates}
In this section, let us try to understand the shape of the B-mode angular power spectrum $C_{\ell}^{BB}$ in the EA theory in an analytic way. The following discussion is similar to the one introduced in \cite{Pritchard:2004qp} for the B-mode spectrum from the inflationary gravitational wave. 


The starting point is the integral solution for the $B_{\ell}$ which was alreadly introduced in Eq.~(\ref{eq:Bl-sol}):
\begin{equation}
B_{\ell}(\eta_{0}) = -\frac{\ell-1}{\ell+1}\int^{\eta_{0}} d\eta \dot{\tau} e^{-\tau} \Psi_{\ell}[k(\eta_{0}-\eta)]\zeta. \nonumber
\end{equation}
Using the approximation for the visibility function, $\dot{\tau}e^{-\tau} \sim \delta(\eta-\eta_{R})$, we see that it is important to know $\zeta$ at the last scattering surface for determining the B-mode polarization.

Now, we expand the multipole equations (\ref{eq:photon-multipole}), (\ref{eq:El-multipole}) and (\ref{eq:Bl-multipole}) in terms of $k/\dot{\tau}$. Neglecting $I_{0}$, $I_{\ell}$
$(\ell \ge 3)$, and $E_{\ell}$
$(\ell \neq 2)$, we have
\begin{eqnarray}
I_{1}^{\prime}+\frac{k}{2}I_{2} &=& -\dot{\tau}\left(I_{1}-\frac{4}{3}v_{b}\right), \\
I_{2}^{\prime}-\frac{2k}{5}I_{1} &=& -\dot{\tau} \left(I_{2}-\frac{2}{15}\zeta\right) +\frac{8}{15}k\sigma, \label{eq:I2}\\
E_{2}^{\prime} &=& -\dot{\tau} \left( E_{2}-\frac{2}{15}\zeta\right) \label{eq:E2}.  
\end{eqnarray} 
At leading order, Eq.~(\ref{eq:E2}) reduces to
\begin{eqnarray}
E_{2} = \frac{2}{15}\zeta \;\;\Leftrightarrow \;\; E_{2} = \frac{1}{16}I_{2}.
\end{eqnarray}
Substituting this into Eq.~(\ref{eq:I2}) and neglecting a higher order term, we find 
\begin{equation}
I_{2} =\frac{32}{75}\frac{k}{\dot{\tau}}\left(I_{1}+\frac{4}{3}\sigma\right)
\end{equation}
and 
\begin{equation}
\zeta = \frac{3}{4}I_{2}-\frac{9}{2}E_{2}=\frac{1}{5}\frac{k}{\dot{\tau}}\left(I_{1}+\frac{4}{3}\sigma\right).
\end{equation}
The initial conditions we have adopted in Sec.~IV imply that at the early times
$I_{1}=4v_{\gamma}/3=q_{\gamma}/\rho_{\gamma}$ and $v_{\nu}$ can be neglected compared with $\sigma$. Then, we have the relation:  
\begin{equation}
\zeta = \frac{4}{15} \frac{k}{\dot{\tau}}\sigma.
\end{equation}
Under the same approximation, $\sigma$ and the fluctuation in the aether field $V$ are related as
\begin{equation}
\sigma=-\frac{c_{13}}{1+c_{13}}V. \label{eq:sigma-v}
\end{equation}
The right-hand side of the evolution equation for the aether field (\ref{per-ae})
is expressed by the aether field itself and this equation becomes the closed form,
\begin{eqnarray}
&&V^{\prime\prime} + 2\cH V^{\prime}+c_{v}^{2}k^{2}V \nonumber \\&&\;\;
+\left[\left(1+\frac{\alpha}{c_{14}}\right)\cH^{2}
-\left(1-\frac{\alpha}{c_{14}}\right)\cH^{\prime}\right]V=0,
\label{eq:aether-evolution}
\end{eqnarray}
where the sound speed $c_{v}$ is given by
\begin{equation}
c_{v}^{2}=\frac{c_{1}}{c_{14}}\left[ 1-\frac{c_{13}^{2}}{2c_{1}(1+c_{13})}\right].
\end{equation} 
Equation (\ref{eq:aether-evolution}) tells us that
on superhorizon scales ($c_{v}k\eta \ll 1$)
$V$ is proportional to $\eta^{\nu}$ (or, equivalently, $S^{\nu}$)
in the radiation-dominated stage and to $\eta^{\nu_{{\rm m}}}$ (or, equivalently, $S^{\nu_{{\rm m}}/2}$)
in the matter-dominated stage, while on subhorizon scales
($c_{v}k\eta \gg 1$) $V$ decays as $S^{-1}$ with oscillations $\cos(kc_{v}\eta)$.

Bearing these facts in mind, we can
derive the wavenumber dependence of $V$ at recombination $\eta_{{\rm rec}}$.
Superhorizon modes evolve in the same way and
keep their dependence on $k$, ${\cal A}_k$, until horizon crossing.
Therefore, the superhorizon modes at
$\eta_{{\rm rec}}$ ($k < 1/c_{v}\eta_{{\rm rec}}$)
keep the primordial $k$-dependence, ${\cal A}_k$.
After horizon crossing, $V$ decays as $S^{-1}$, so that
$V(\eta_{\rm rec})/V(\eta_{*})\propto S(\eta_{*})$, where $\eta_*:=1/c_vk$.
The modes with $ 1/c_{v}\eta_{{\rm rec}} < k < 1/c_{v}\eta_{{\rm eq}}$
reenter the horizon in the matter-dominated stage,
where $\eta_{\rm eq}$ referes to the radiation-matter equality time.
For these modes, we have
$S(\eta_{*})\propto \eta_{*}^{2} \propto k^{-2}$ and
$V(\eta_{*})\propto \eta_{*}^{\nu_{\rm m}}\propto k^{-\nu_{\rm m}}$,
and hence
$V(\eta_{\rm rec}) \propto k^{-2-\nu_{\rm m}}{\cal A}_{k}$.
Similarly, the modes with
$1/c_{v}\eta_{{\rm eq}} < k$ reenter the horizon in the radiation-dominated stage
and so $V(\eta_{\rm rec})\propto k^{-1-\nu}{\cal A}_{k}$.






In summary, the wavenumber dependence of $V(\eta_{\rm rec})$ is:
\begin{equation}
V(\eta_{\rm rec})\propto 
\begin{cases}
 {\cal A}_{k} & (k < 1/c_{v}\eta_{{\rm rec}}), \\
 k^{-2-\nu_{{\rm m}}}{\cal A}_{k} & (1/c_{v}\eta_{{\rm rec}} < k < 1/c_{v}\eta_{{\rm eq}}), \\
 k^{-1-\nu}{\cal A}_{k} & (1/c_{v}\eta_{{\rm eq}} < k).
\end{cases}
\end{equation}
In the above argument
we ignored the subhorizon oscillation of $V$.

In all the above calculations, we have neglected the matter velocities $v_{i}$ compared with $\sigma$. Our numerical results in Fig.~\ref{fig:params1} justify
the approximation. The amplitudes of $V$ and $\sigma$ are
large enough to neglect $v_{i}$ long before the recombination epoch. 
In Fig.~\ref{fig:params2}, we can see that the evolution of $\sigma$ tracks that of $V$ and the ratio is approximately equal to $-c_{13}/(1+c_{13})$ until
the neutrino velocity $v_{\nu}$ becomes comparable.
(In Fig.~\ref{fig:params2}, $c_{13}=-0.3$, so that the ratio is $3/7$.) 


We are now
in position to derive the simple scaling relation for $C_{\ell}^{BB}$ in an analytic way. The CMB B-mode power spectrum is roughly expressed as
\begin{equation}
C_{\ell}^{BB}\sim \int \D\ln k\,{\cal P}_{V}(k) B_{\ell} B_{\ell} ,
\end{equation}
where ${\cal P}_{V}(k)$ is the dimensionless primordial power
spectrum for $V$. From the above discussion, we see
\begin{eqnarray}
B_{\ell} (\eta_{0}) &\sim& \int^{\eta_{0}} \D\eta \delta(\eta-\eta_{{\rm rec}}) \frac{\ell j_{\ell}[k(\eta_{0}-\eta)]}{k(\eta_{0}-\eta)} \frac{k}{\dot{\tau}}\frac{V}{{\cal A}_{k}}\frac{c_{13}}{1+c_{13}} \nonumber \\
&\sim& \frac{\ell j_{\ell}[k(\eta_{0}-\eta_{{\rm rec}})]}{k(\eta_{0}-\eta_{{\rm rec}})}\frac{k}{\dot{\tau}}\frac{V(\eta_{{\rm rec}})}{{\cal A}_{k}} \frac{c_{13}}{1+c_{13}}. 
\end{eqnarray}
Using the fact that the projection factor $\ell j_{\ell}(x)/x$ has a peak at $\ell \sim x$, the power spectrum
reduces to
\begin{equation}
C_{\ell}^{BB} \sim \left(\frac{V}{{\cal A}_k}\right)^{2}_{k=\frac{\ell}{\eta_{0}-\eta_{\rm rec}}}
\int k^{2+n_v}[\Psi_{\ell}(k(\eta_{0}-\eta_{\rm rec}))]^{2} d\ln k.
\end{equation}
For example, taking $n_v=1$,
the above integral 
can be evaluated to give 
the scaling 
\begin{equation}
\ell(\ell+1)C_{\ell}^{BB}\propto
\begin{cases}
\ell^{3} & (\ell < \ell_{\rm rec}), \\
\ell^{-1-2\nu_{{\rm m}}} & (\ell_{\rm rec}< \ell < \ell_{\rm eq}), \\
\ell^{1-2\nu} & (\ell_{\rm eq}<\ell < \ell_{\Delta \eta_{{\rm rec}}}), \\
\end{cases}
\label{cl_scaling}
\end{equation}
showing a peak at $\ell_{\rm peak}\sim\ell_{\rm rec}$, where we have defined $\ell_{\rm rec}:=(\eta_0-\eta_{\rm rec})/c_{v}\eta_{\rm rec}$ and $\ell_{\rm eq}:=(\eta_0-\eta_{\rm rec})/c_{v}\eta_{\rm eq}$. $\ell_{\Delta \eta_{{\rm rec}}}$ represents the scale over which the
phase-damping effect shows up due to the width of the last-scattering surface. On scales $\ell > \ell_{\Delta \eta_{{\rm rec}}}$, the above approximation is no longer justified.
In deriving the scaling we used the following integral formula for the Bessel function,
\begin{eqnarray}
&&\int \D \ln k\, k^{2+m}\Psi_l(k) \nonumber \\&&\;\;
=\ell^{2}\frac{\sqrt{\pi}}{4}
\frac{\Gamma(1-m/2)}{\Gamma (3/2-m/2)}\frac{\Gamma(l+m/2)}{\Gamma(l+2-m/2)}.
\end{eqnarray}

\begin{figure}[t]
\includegraphics[keepaspectratio=true,width=8cm]{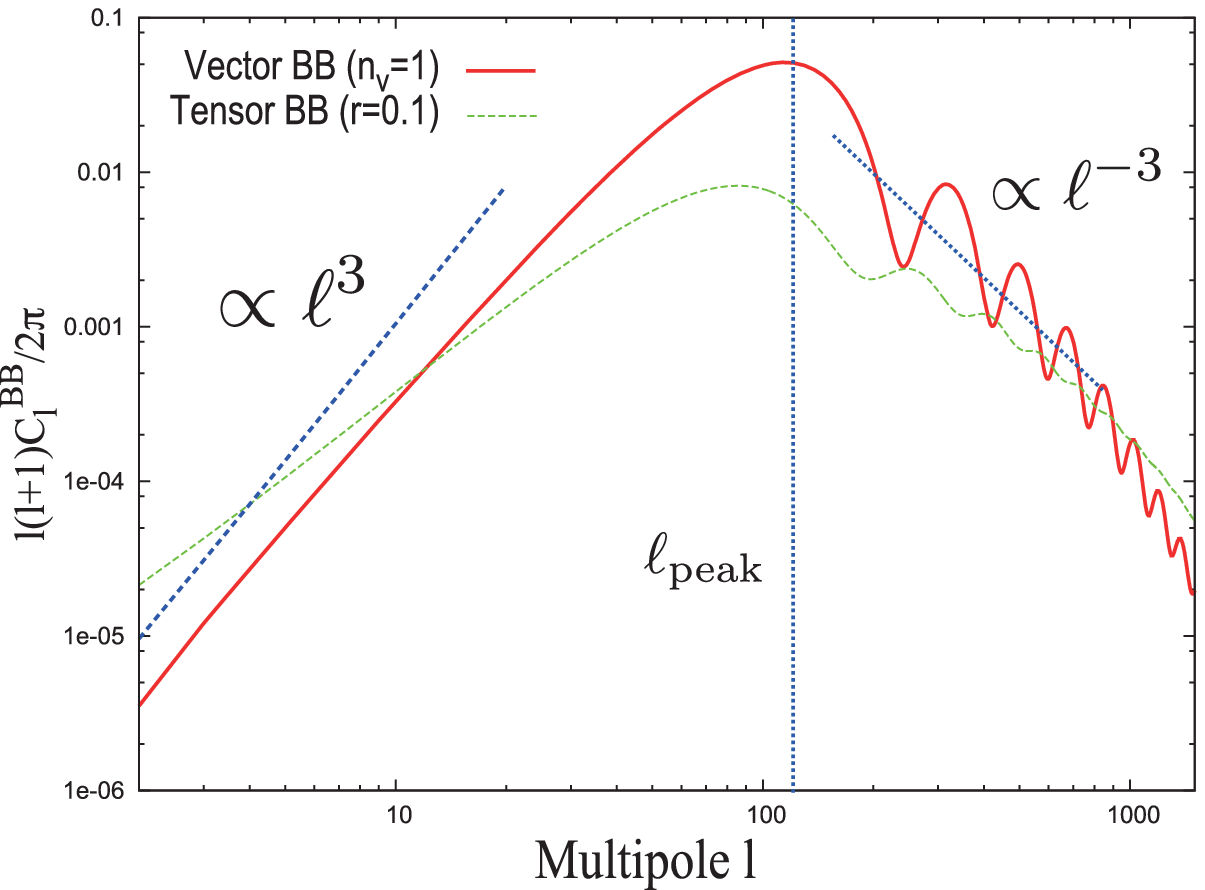}
\caption{(a) Scaling for the illustrative case with $n_v=1$ and $\alpha=-c_{14}=0.2$.
The other parameters are given by $c_{13}=-0.3$, and $c_1=-0.1$;
(b) Parameter dependence of the spectrum.
In the two examples $c_{14}$ is different while the other parameters
are fixed as $n_v=1$, $c_{13}=-0.3$, and $c_1=-0.1$. The primordial amplitudes are arbitrary.}
\label{fig:P48_ex2}
\end{figure}

The scaling behavior can indeed be seen in Fig.~\ref{fig:P48_ex2} for
the illustrative case $\alpha=-c_{14}\ (\nu=\nu_{\rm m}=1)$, though
the scaling in the range
$\ell_{\rm eq}<\ell<\ell_{\Delta \eta_{{\rm rec}}}$ is not clearly seen
due to the phase-damping effect~\cite{Pritchard:2004qp}. Actually, in order to evaluate the shape at the smallest scales correctly,
we must take into account more complicated physics such as neutrino anisotropic stresses and the Silk-damping effect.
We would emphasize that
our numerical calculations incorporate
all of these effects. 
For instance, in Fig.~\ref{fig:P48_ex2}
we can confirm the oscillatory behavior
at $\ell > \ell_{{\rm peak}}$ arising from the subhorizon oscillation of $V$.
(In Fig~\ref{fig:P48_ex2},
the B-mode spectrum from tensor perturbations in GR is also plotted for comparison.)

\begin{figure}[t]
\includegraphics[keepaspectratio=true,width=8cm]{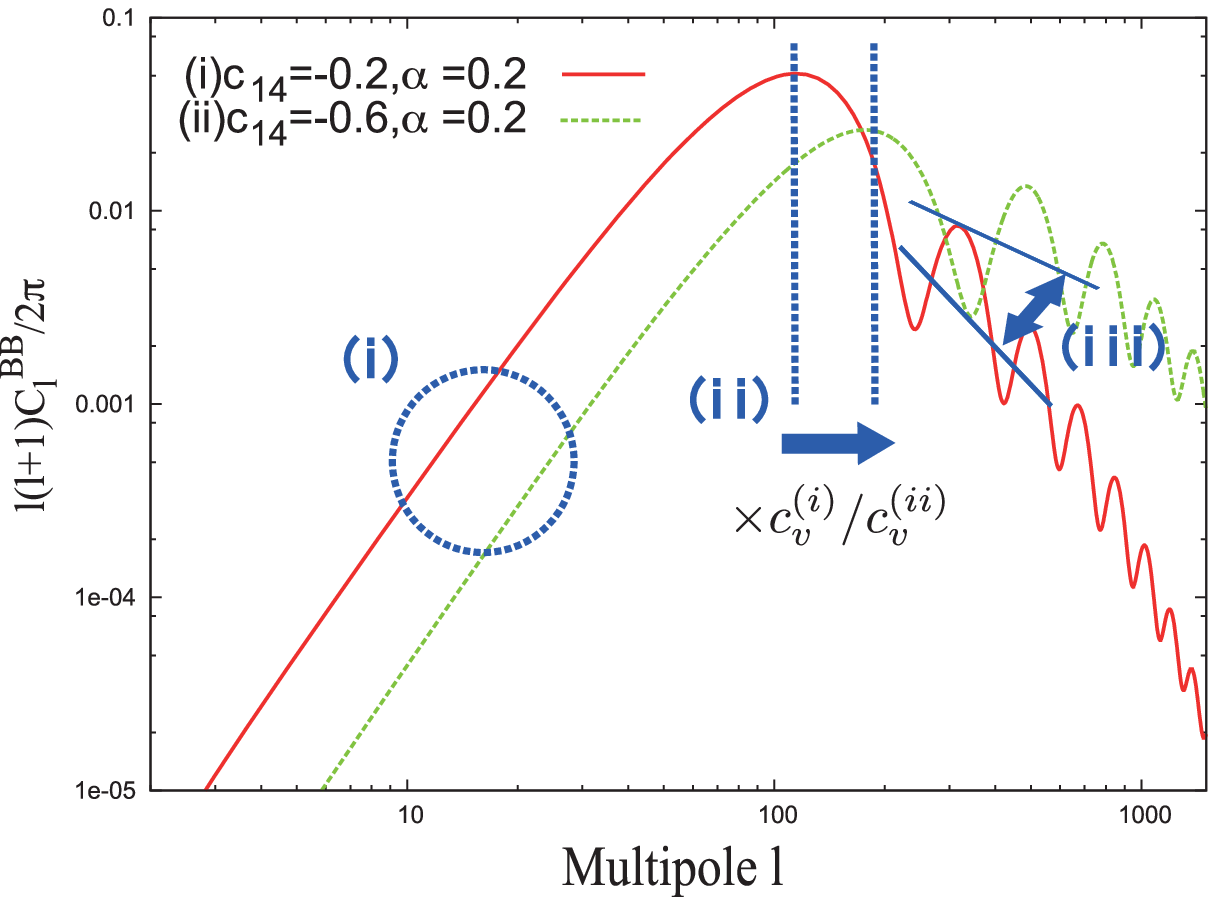}
\caption{(a) Scaling for the illustrative case with $n_v=1$ and $\alpha=-c_{14}=0.2$.
The other parameters are given by $c_{13}=-0.3$, and $c_1=-0.1$;
(b) Parameter dependence of the spectrum.
In the two examples $c_{14}$ is different while the other parameters
are fixed as $n_v=1$, $c_{13}=-0.3$, and $c_1=-0.1$. The primordial amplitudes are arbitrary.}
\label{fig:param_dep}
\end{figure}

We can also gain an understanding of how the shape of the angular power spectrum depends on the model parameters. From Fig.~\ref{fig:param_dep}, one can confirm the following three things: (i) Since the angular power spectrum on the largest scales $\ell < \ell_{{\rm rec}}$ depends only on the primordial spectrum, the plotted examples show the same scaling at this scale. This means that the observation of the polarization at the largest scale can determine the primordial spectral index; (ii) The peak position is inversely proportional to the sound velocity of the aether vector perturbation $c_{v}$. Actually, if we find the peak of the observed B-mode spectrum, the information is directly converted into the exact value of $c_{v}$ in case the other cosmological parameters are already determined; (iii) The difference of the small scale scaling arises due to the difference of the growth rate of $V$ on superhorizon scales, which can be seen in Eq. (\ref{cl_scaling}).

\section{Summary}

In this paper, we have considered the Einstein-aether theory
in which gravity is modified through the additional vector degree of freedom (the aether),
and studied
possible signatures of the aether field in the CMB polarization.
In standard GR, vector cosmological perturbations simply decay without sources,
and hence they are less relevant to observations.
In the Einstein-aether theory, however, the vector modes are
dynamical, so that they could leave imprints on the CMB signals
in a way similar to the inflationary gravitational waves.

Using the CAMB code modified to incorporate the aether vector perturbation, we have computed the CMB B-mode polarization power spectrum. 
We have found that the amplitude of the B-mode polarization from the aether vector 
can be larger than that from inflationary gravitational waves with $r={\cal O}(0.1)$
on small angular scales, which would be measurable with near future CMB probes. 

Moreover, we have found that,
for a set of parameters $c_{i}$ evading all the existing observational and experimental
constraints,
the B-mode polarization spectrum indeed shows distinguishable features
from that from inflationary gravitational waves in GR.
Thus,
the B-mode polarization spectrum would potentially be a crucial test for the Einstein-aether theory.    

We
performed analytical calculations, by which we clarified
the physical origins of the shape of the B-mode power spectrum.
The scaling relations derived in an analytical way
turned out to be consistent with
the numerical results.

In this paper, we have simply assumed
the primordial amplitude of the vector perturbation
which is generated anyway in the presence of the aether field.
The primordial spectrum actually depends upon
the underlying inflation model and the reheating process.
It would be interesting to explore whether
the B-mode signal is measurable or not
assuming concrete inflation models.
We will report elsewhere
the evolution of vector perturbation
during the whole history of the Universe
starting from inflation in the EA theory.

\acknowledgments
This work was partially supported by
JSPS Grant-in-Aid for Research Activity Start-up No.~22840011 (T.K.).
M.N. is supported bye JSPS through research fellowship.

\appendix

\section{Observational constraint}
In this appendix, we summarize the existing observational constraints on the aether parameters $c_{1},\cdots,c_{4}$, following \cite{Garriga, Jacobson:2008aj}.

\subsection{Post-Newtonian limits}
Parametrized post-Newtonian (PPN) parameters in the EA theory have already been analyzed in \cite{Eling:2003rd,Foster:2005dk}. Two PPN parameters, the Eddington-Robertson-Schiff parameters $\beta$ and $\gamma$, are identical to those in pure GR~\cite{Eling:2003rd}. The Whitehead parameter, $\xi$, which characterizes a peculiar sort of three-body interaction vanishes in the EA theory~\cite{Foster:2005dk}, and the five energy-momentum conservation parameters $\alpha_{3}$ and $\zeta_{1,2,3,4}$ vanish because the theory is derived from a Lagrangian.

The aether defines a preferred frame, and its effect is encoded in the remained PPN parameters $\alpha_{1}$ and $\alpha_{2}$. The exact values of $\alpha_{1}$ and $\alpha_{2}$ are found in \cite{Foster:2005dk}:
\begin{eqnarray}
\alpha_{1} &=& \frac{-8(c_{3}^{2}+c_{1}c_{4})}{2c_{1}-c_{1}^{2}+c_{3}^{2}}, \\
\alpha_{2} &=& \frac{\alpha_{1}}{2}-\frac{(2c_{13}-c_{14})(\alpha+c_{14})}{c_{123}(2-c_{14})}.
\end{eqnarray}
The easiest way to pass the stringent observational constraints, $\alpha_{1}\lesssim 10^{-4}$ and $\alpha_{2} \lesssim 4 \times 10^{-7}$~\cite{Will:2005va}, is to set $\alpha_{1,2}$ exactly to zero by imposing the conditions
\begin{eqnarray}
c_{2} &=& \frac{-2c_{1}^{2}-c_{1}c_{3}+c_{3}^{2}}{3c_{1}}, \label{PPN-vanish1} \\
c_{4} &=& -\frac{c_{3}^{2}}{c_{1}} \label{PPN-vanish2},
\end{eqnarray}
which is possible since EA theory has four free parameters $c_{i} (i=1-4)$. Under these conditions, all the PPN parameters in the EA theory coincide with those of GR.



\subsection{Stability of each perturbation mode}
Linear perturbations around the flat or FRW metric have been studied in~\cite{Garriga,Jacobson:2004ts}. They concluded that quantum and classical stabilities of tensors (spin-2 modes), vectors (spin-1), and scalars (spin-0) constrain the aether parameters to be
\begin{gather}
c_{13} > -1, \\
-2 \le c_{14} < 0, \quad c_{123} < 0, \\
2c_{1} \le c_{13}^{2}(1+c_{13}). 
\end{gather}

\subsection{Radiation damping and strong self-field effects}
The radiation damping rate in the weak field limit was first calculated in~\cite{Foster:2006az}, and then strong field effects are included in~\cite{Foster:2007gr}. They found that in general the damping rate in the EA theory is different from that of GR. However, in the special case with $\alpha_{1}=\alpha_{2}=0$, it is identical provided that the quadrupole coefficient of the radiation damping rate,
\begin{eqnarray}
&&{\cal A} = \left(1+\frac{c_{14}}{2}\right)\bigg[\frac{1}{c_{t}}-\frac{2c_{14}c_{13}^{2}}{(2c_{1}+c_{13}c_{-})^{2}}\frac{1}{c_{v}} \nonumber \\&&\:\:
-\frac{c_{14}}{6(2+c_{14})}\left(3+\frac{2\alpha_{2}-\alpha_{1}}{2(2c_{13}-c_{14})}\right)^{2}\frac{1}{c_{s}}\bigg],
\end{eqnarray}
is equal to one. Here we defined the new parameters 
\begin{eqnarray}
c_{t}^{2} &:=& \frac{1}{1+c_{13}}, \\
c_{s}^{2} &:=& \frac{(2+c_{14})c_{123}}{(1+c_{13})(2-\alpha)c_{14}},
\end{eqnarray}
which correspond to the velocity of tensor modes and scalar mode, respectively, and $c_{-}:= c_{1}-c_{3}$. 

Foster~\cite{Foster:2007gr} derived several constraints
by considering strong field effects. All of them reduce to the condition $|c_{i}| \lesssim {\cal O}(0.1)$ with $\alpha_{1}=\alpha_{2}=0$ under the current observational uncertainties.


\subsection{Cherenkov Radiation}
If the sound speed of each perturbation mode were smaller than the speed of light, then the ultra-high-energy particles would loose their energies into the mode in a similar way to Cherenkov radiation.
Using this fact, Elliott {\it et al}.~\cite{Elliott:2005va} derived limits on the aether parameters. If all the aether modes propagate superluminally, we do not need to take into account their constraints.

The conditions for superluminal propagation of tensors, vectors, and scalars are 
\begin{gather}
c_{13} \le 0, \\
(2+c_{14})c_{123} \le (2-\alpha)(1+c_{13})c_{14}, \\
 2c_{4} \ge -c_{13}^{2}/(1+c_{13}),
\end{gather}
respectively. The connection between superluminality and violation of causality has been under debates~\cite{Adams:2006sv,Bonvin:2006vc,Ellis:2007ic,Babichev:2007dw,Gorini:2007ta}. Since the above conditions together with the PPN vanishing conditions (\ref{PPN-vanish1}) and (\ref{PPN-vanish2}) are equivalent to the stability condition of scalar and vector modes, we here allow for the superluminal propagation in the EA theory.

\subsection{Scalar mode constraint}
Armendariz-Picon {\it et al}.~\cite{Garriga} have formulated the evolution equations for perturbations around the FRW background metric and calculated CMB temperature anisotropy spectrum. They found mainly two constraints on the aether parameters:
\begin{itemize}
\item In order for the scalar isocurvature mode not to grow on superhorizon scales,
\begin{equation}
\alpha \le -c_{14} \label{growing-isocurvature}
\end{equation}
must be satisfied.   
\item In order not to have too large an anisotropic stress,
\begin{equation}
|c_{13}| \lesssim 1 \label{scalar-isocurvature-cond}
\end{equation}
must be satisfied.
\end{itemize} 

\section{Allowed parameter values}
If one imposes that the PPN parameters coincide {\em exactly} with those in GR, one has two constraints derived from $\alpha_{1} = \alpha_{2} =0$. We are now left with two free parameters, for which we use $c_{13}=c_{1}+c_{3}$ and $c_{-} = c_{1}-c_{3}$.
Interestingly, the special combination $\alpha = -c_{14}$ is automatically satisfied, and we have 
\begin{eqnarray}
\alpha = -c_{14} = -2 \frac{c_{13}c_{-}}{c_{13}+c_{-}}.
\end{eqnarray}
Imposing that the perturbations are stable and superluminal, we can restrict the two parameters within the range as
\begin{eqnarray}
-1 \le &c_{13}& \le 0, \label{c13region}\\
\frac{c_{13}}{3(1+c_{13})} \le &c_{-}& \le 0. \label{c-region}  
\end{eqnarray}
For the parameters satisfying the above constraints, we have safely a cosmological solution (see Eq. (\ref{background_stability})), and we we do not have a growing isocurvature mode (see Eq. (\ref{growing-isocurvature})).

Imposing further that ${\cal A}=1$, one can evade the constraint from radiation damping rate. Thus, we are finally left with a single parameter, say, $c_{13}$, within the range (\ref{c13region})~\cite{Foster:2007gr}.
Taking $|c_{i}| \lesssim {\cal O}(0.1)$, we choose to use 
\begin{eqnarray}
&& c_{1} =-0.019,\; c_{2} = 0.014,\;
c_{3}=-0.011, \;c_{4}=0.0063, \nonumber \\&&
c_{v}= 1.241, \;\alpha=-c_{14}=0.0128
\end{eqnarray}
in Sec.~V of the main text.



\begin{thebibliography}{99}

\bibitem{bd}
Y.~Fujii, K.~Maeda,
  ``The scalar-tensor theory of gravitation,''
  Cambridge, USA: Univ. Pr. (2003) 240 p.

\bibitem{cham}
D.~F.~Mota and J.~D.~Barrow,
  Phys.\ Lett.\  B {\bf 581}, 141 (2004)
  [arXiv:astro-ph/0306047];
J.~Khoury and A.~Weltman,
  Phys.\ Rev.\ Lett.\  {\bf 93}, 171104 (2004)
  [arXiv:astro-ph/0309300];
J.~Khoury and A.~Weltman,
  Phys.\ Rev.\  D {\bf 69}, 044026 (2004)
  [arXiv:astro-ph/0309411].

\bibitem{fr}
A.~A.~Starobinsky,
  JETP Lett.\  {\bf 86}, 157-163 (2007).
  [arXiv:0706.2041 [astro-ph]];
  W.~Hu, I.~Sawicki,
  Phys.\ Rev.\  {\bf D76}, 064004 (2007).
  [arXiv:0705.1158 [astro-ph]];
  S.~A.~Appleby, R.~A.~Battye,
  Phys.\ Lett.\  {\bf B654}, 7-12 (2007).
  [arXiv:0705.3199 [astro-ph]];
S.~Nojiri and S.~D.~Odintsov,
  Phys.\ Rev.\  D {\bf 77}, 026007 (2008)
  [arXiv:0710.1738 [hep-th]].

\bibitem{gal}
A.~Nicolis, R.~Rattazzi and E.~Trincherini,
  Phys.\ Rev.\  D {\bf 79}, 064036 (2009)
  [arXiv:0811.2197 [hep-th]];
  C.~Deffayet, G.~Esposito-Farese and A.~Vikman,
  Phys.\ Rev.\  D {\bf 79}, 084003 (2009)
  [arXiv:0901.1314 [hep-th]];
  C.~Deffayet, S.~Deser and G.~Esposito-Farese,
  Phys.\ Rev.\  D {\bf 80}, 064015 (2009)
  [arXiv:0906.1967 [gr-qc]].


\bibitem{gde}
  N.~Chow and J.~Khoury,
  Phys.\ Rev.\  D {\bf 80}, 024037 (2009)
  [arXiv:0905.1325 [hep-th]];
  F.~P.~Silva and K.~Koyama,
  Phys.\ Rev.\  D {\bf 80}, 121301 (2009)
  [arXiv:0909.4538 [astro-ph.CO]];
    T.~Kobayashi, H.~Tashiro and D.~Suzuki,
  Phys.\ Rev.\  D {\bf 81}, 063513 (2010)
  [arXiv:0912.4641 [astro-ph.CO]];
  T.~Kobayashi,
  Phys.\ Rev.\  D {\bf 81}, 103533 (2010)
  [arXiv:1003.3281 [astro-ph.CO]];
  R.~Gannouji and M.~Sami,
  Phys.\ Rev.\  D {\bf 82}, 024011 (2010)
  [arXiv:1004.2808 [gr-qc]];
A.~De Felice, S.~Tsujikawa,
  JCAP {\bf 1007}, 024 (2010).
  [arXiv:1005.0868 [astro-ph.CO]];
A.~De Felice, S.~Mukohyama, S.~Tsujikawa,
  Phys.\ Rev.\  {\bf D82}, 023524 (2010).
  [arXiv:1006.0281 [astro-ph.CO]];
A.~De Felice, S.~Tsujikawa,
  Phys.\ Rev.\ Lett.\  {\bf 105}, 111301 (2010).
  [arXiv:1007.2700 [astro-ph.CO]];
  A.~Ali, R.~Gannouji, M.~Sami,
  Phys.\ Rev.\  {\bf D82}, 103015 (2010).
  [arXiv:1008.1588 [astro-ph.CO]];
  A.~De Felice, S.~Tsujikawa,
  [arXiv:1008.4236 [hep-th]];
  S.~Nesseris, A.~De Felice, S.~Tsujikawa,
  Phys.\ Rev.\  {\bf D82}, 124054 (2010).
  [arXiv:1010.0407 [astro-ph.CO]];
  A.~De Felice, R.~Kase, S.~Tsujikawa,
  Phys.\ Rev.\  {\bf D83}, 043515 (2011).
  [arXiv:1011.6132 [astro-ph.CO]].


\bibitem{tests}
See, {\em e.g.}, B.~Jain, J.~Khoury,
  Annals Phys.\  {\bf 325}, 1479-1516 (2010).
  [arXiv:1004.3294 [astro-ph.CO]].

\bibitem{massiveg}
N.~Arkani-Hamed, H.~Georgi and M.~D.~Schwartz,
  Annals Phys.\  {\bf 305}, 96 (2003)
  [arXiv:hep-th/0210184];
S.~L.~Dubovsky,
  JHEP {\bf 0410}, 076 (2004)
  [arXiv:hep-th/0409124];
 D.~Blas, D.~Comelli, F.~Nesti and L.~Pilo,
  Phys.\ Rev.\  D {\bf 80}, 044025 (2009)
  [arXiv:0905.1699 [hep-th]];
 A.~H.~Chamseddine and V.~Mukhanov,
  JHEP {\bf 1008}, 011 (2010)
  [arXiv:1002.3877 [hep-th]];
V.~A.~Rubakov,
  arXiv:hep-th/0407104.

\bibitem{bigra}
T.~Damour and I.~I.~Kogan,
  Phys.\ Rev.\  D {\bf 66}, 104024 (2002)
  [arXiv:hep-th/0206042].

\bibitem{Jacobson:2000xp}
  T.~Jacobson and D.~Mattingly,
  Phys.\ Rev.\  D {\bf 64}, 024028 (2001)
  [arXiv:gr-qc/0007031].


\bibitem{Carroll:2004ai}
  S.~M.~Carroll and E.~A.~Lim,
  Phys.\ Rev.\  D {\bf 70} (2004) 123525
  [arXiv:hep-th/0407149].

\bibitem{Lim:2004js}
  E.~A.~Lim,
  Phys.\ Rev.\  D {\bf 71}, 063504 (2005)
  [arXiv:astro-ph/0407437].

\bibitem{Li:2007xw}
  B.~Li, J.~D.~Barrow and D.~F.~Mota,
  Phys.\ Rev.\  D {\bf 76}, 104047 (2007)
  [arXiv:0707.2664 [gr-qc]].

\bibitem{Zuntz:2008zz}
  J.~A.~Zuntz, P.~G.~Ferreira and T.~G.~Zlosnik,
  Phys.\ Rev.\ Lett.\  {\bf 101}, 261102 (2008)
  [arXiv:0808.1824 [gr-qc]].

\bibitem{Zuntz:2010jp}
  J.~Zuntz, T.~G.~Zlosnik, F.~Bourliot, P.~G.~Ferreira and G.~D.~Starkman,
  Phys.\ Rev.\  D {\bf 81}, 104015 (2010)
  [arXiv:1002.0849 [astro-ph.CO]].


\bibitem{Garriga}
  C.~Armendariz-Picon, N.~F.~Sierra and J.~Garriga,
  JCAP {\bf 1007}, 010 (2010)
  [arXiv:1003.1283 [astro-ph.CO]].

\bibitem{Eling:2006df}
  C.~Eling and T.~Jacobson,
  Class.\ Quant.\ Grav.\  {\bf 23} (2006) 5625
  [Erratum-ibid.\  {\bf 27} (2010) 049801]
  [arXiv:gr-qc/0603058].



\bibitem{Eling:2007xh}
  C.~Eling, T.~Jacobson and M.~Coleman Miller,
  Phys.\ Rev.\  D {\bf 76} (2007) 042003
  [Erratum-ibid.\  D {\bf 80} (2009) 129906]
  [arXiv:0705.1565 [gr-qc]].


\bibitem{Eling:2006ec}
  C.~Eling and T.~Jacobson,
  Class.\ Quant.\ Grav.\  {\bf 23} (2006) 5643
  [Erratum-ibid.\  {\bf 27} (2010) 049802]
  [arXiv:gr-qc/0604088].

\bibitem{Tamaki:2007kz}
  T.~Tamaki, U.~Miyamoto,
  Phys.\ Rev.\  {\bf D77}, 024026 (2008).
  [arXiv:0709.1011 [gr-qc]].


\bibitem{blas}
  D.~Blas, O.~Pujolas and S.~Sibiryakov,
  Phys.\ Rev.\ Lett.\  {\bf 104}, 181302 (2010)
  [arXiv:0909.3525 [hep-th]].


\bibitem{horava}
  P.~Horava,
  Phys.\ Rev.\  D {\bf 79}, 084008 (2009)
  [arXiv:0901.3775 [hep-th]].

\bibitem{Jacobson:2010mx}
  T.~Jacobson,
  Phys.\ Rev.\  D {\bf 81}, 101502 (2010)
  [Erratum-ibid.\  D {\bf 82}, 129901 (2010)]
  [arXiv:1001.4823 [hep-th]].

\bibitem{cphealthy}
 T.~Kobayashi, Y.~Urakawa, M.~Yamaguchi,
  JCAP {\bf 1004}, 025 (2010);
  A.~Cerioni, R.~H.~Brandenberger,
  [arXiv:1008.3589 [hep-th]].






\bibitem{Hu:1997hp}
  W.~Hu and M.~J.~White,
  Phys.\ Rev.\  D {\bf 56} (1997) 596
  [arXiv:astro-ph/9702170].

\bibitem{Pen:1997ae}
  U.~L.~Pen, U.~Seljak and N.~Turok,
  Phys.\ Rev.\ Lett.\  {\bf 79}, 1611 (1997)
  [arXiv:astro-ph/9704165];
  U.~Seljak, U.~L.~Pen and N.~Turok,
  Phys.\ Rev.\ Lett.\  {\bf 79}, 1615 (1997)
  [arXiv:astro-ph/9704231];
  N.~Turok, U.~L.~Pen and U.~Seljak,
  Phys.\ Rev.\  D {\bf 58}, 023506 (1998)
  [arXiv:astro-ph/9706250].

\bibitem{Seljak:2006hi}
  U.~Seljak and A.~Slosar,
  Phys.\ Rev.\  D {\bf 74}, 063523 (2006)
  [arXiv:astro-ph/0604143].


\bibitem{Starobinsky:1979ty}
  A.~A.~Starobinsky,
  JETP Lett.\  {\bf 30}, 682 (1979)
  [Pisma Zh.\ Eksp.\ Teor.\ Fiz.\  {\bf 30}, 719 (1979)].


\bibitem{Lewis-Vector}
  A.~Lewis,
  Phys.\ Rev.\  D {\bf 70}, 043518 (2004)
  [arXiv:astro-ph/0403583].


\bibitem{2ndvec1}
K.~Ichiki, K.~Takahashi, H.~Ohno, H.~Hanayama, N.~Sugiyama,
  Science {\bf 311}, 827-829 (2006).
  [astro-ph/0603631].

\bibitem{2ndvec2}
T.~Kobayashi, R.~Maartens, T.~Shiromizu, K.~Takahashi,
  Phys.\ Rev.\  {\bf D75}, 103501 (2007);
  S.~Maeda, S.~Kitagawa, T.~Kobayashi, T.~Shiromizu,
  Class.\ Quant.\ Grav.\  {\bf 26}, 135014 (2009).
  [arXiv:0805.0169 [astro-ph]].

\bibitem{2ndvec3}
A.~J.~Christopherson, K.~A.~Malik, D.~R.~Matravers,
  Phys.\ Rev.\  {\bf D79}, 123523(2009).
  [arXiv:0904.0940 [astro-ph.CO]];
  A.~J.~Christopherson, K.~A.~Malik, D.~R.~Matravers,
  [arXiv:1008.4866 [astro-ph.CO]].

\bibitem{2ndvec4}
T.~H.~-C.~Lu, K.~Ananda, C.~Clarkson,
  Phys.\ Rev.\  {\bf D77}, 043523 (2008).
  [arXiv:0709.1619 [astro-ph]];
  T.~H.~-C.~Lu, K.~Ananda, C.~Clarkson, R.~Maartens,
  JCAP {\bf 0902}, 023 (2009).
  [arXiv:0812.1349 [astro-ph]].

\bibitem{Pitrou:2008hy}
  C.~Pitrou,
  Class.\ Quant.\ Grav.\  {\bf 26}, 065006 (2009).
  [arXiv:0809.3036 [gr-qc]];
  C.~Pitrou,
  Gen.\ Rel.\ Grav.\  {\bf 41}, 2587 (2009)
  [arXiv:0809.3245 [astro-ph]].


\bibitem{Lewis:2004ef}
  A.~Lewis,
  Phys.\ Rev.\  D {\bf 70} (2004) 043011
  [arXiv:astro-ph/0406096];
  D.~Paoletti, F.~Finelli and F.~Paci,
  Mon.\ Not.\ Roy.\ Astron.\ Soc.\  {\bf 396}, 523 (2009)
  [arXiv:0811.0230 [astro-ph]];
  J.~R.~Shaw and A.~Lewis,
  Phys.\ Rev.\  D {\bf 81}, 043517 (2010)
  [arXiv:0911.2714 [astro-ph.CO]].


\bibitem{Challinor:1998xk}
  A.~Challinor, A.~Lasenby,
  Astrophys.\ J.\  {\bf 513}, 1-22 (1999).
  [astro-ph/9804301].


\bibitem{Challinor:2000as}
  A.~Challinor,
  Phys.\ Rev.\  {\bf D62}, 043004 (2000).
  [astro-ph/9911481].

\bibitem{Lewis:1999bs}
  A.~Lewis, A.~Challinor and A.~Lasenby,
  Astrophys.\ J.\  {\bf 538} (2000) 473
  [arXiv:astro-ph/9911177].

\bibitem{:2006uk}
    [Planck Collaboration],
  arXiv:astro-ph/0604069;
  D.~Baumann {\it et al.}  [CMBPol Study Team Collaboration],
  AIP Conf.\ Proc.\  {\bf 1141}, 10 (2009)
  [arXiv:0811.3919 [astro-ph]].


\bibitem{Pritchard:2004qp}
  J.~R.~Pritchard and M.~Kamionkowski,
  Annals Phys.\  {\bf 318} (2005) 2
  [arXiv:astro-ph/0412581].




\bibitem{Jacobson:2008aj}
  T.~Jacobson,
  PoS {\bf QG-PH}, 020 (2007)
  [arXiv:0801.1547 [gr-qc]].


\bibitem{Eling:2003rd}
  C.~Eling and T.~Jacobson,
  Phys.\ Rev.\  D {\bf 69} (2004) 064005
  [arXiv:gr-qc/0310044].

\bibitem{Foster:2005dk}
  B.~Z.~Foster and T.~Jacobson,
  Phys.\ Rev.\  D {\bf 73} (2006) 064015
  [arXiv:gr-qc/0509083].

\bibitem{Will:2005va}
  C.~M.~Will,
  Living Rev.\ Rel.\  {\bf 9} (2005) 3
  [arXiv:gr-qc/0510072].



\bibitem{Jacobson:2004ts}
  T.~Jacobson and D.~Mattingly,
  Phys.\ Rev.\  D {\bf 70} (2004) 024003
  [arXiv:gr-qc/0402005].

\bibitem{Foster:2006az}
  B.~Z.~Foster,
  Phys.\ Rev.\  D {\bf 73} (2006) 104012
  [Erratum-ibid.\  D {\bf 75} (2007) 129904]
  [arXiv:gr-qc/0602004].




\bibitem{Foster:2007gr}
  B.~Z.~Foster,
  Phys.\ Rev.\  D {\bf 76} (2007) 084033
  [arXiv:0706.0704 [gr-qc]].

\bibitem{Elliott:2005va}
  J.~W.~Elliott, G.~D.~Moore and H.~Stoica,
  JHEP {\bf 0508} (2005) 066
  [arXiv:hep-ph/0505211].

\bibitem{Adams:2006sv}
  A.~Adams, N.~Arkani-Hamed, S.~Dubovsky, A.~Nicolis and R.~Rattazzi,
  JHEP {\bf 0610} (2006) 014
  [arXiv:hep-th/0602178].

\bibitem{Bonvin:2006vc}
  C.~Bonvin, C.~Caprini and R.~Durrer,
  Phys.\ Rev.\ Lett.\  {\bf 97}, 081303 (2006)
  [arXiv:astro-ph/0606584].

\bibitem{Ellis:2007ic}
  G.~Ellis, R.~Maartens and M.~A.~H.~MacCallum,
  Gen.\ Rel.\ Grav.\  {\bf 39}, 1651 (2007)
  [arXiv:gr-qc/0703121].

\bibitem{Babichev:2007dw}
  E.~Babichev, V.~Mukhanov and A.~Vikman,
  JHEP {\bf 0802}, 101 (2008)
  [arXiv:0708.0561 [hep-th]].

\bibitem{Gorini:2007ta}
  V.~Gorini, A.~Y.~Kamenshchik, U.~Moschella, O.~F.~Piattella and A.~A.~Starobinsky,
  JCAP {\bf 0802}, 016 (2008)
  [arXiv:0711.4242 [astro-ph]].




\bibitem{Graesser:2005bg}
  M.~L.~Graesser, A.~Jenkins and M.~B.~Wise,
  Phys.\ Lett.\  B {\bf 613} (2005) 5
  [arXiv:hep-th/0501223].






\end{thebibliography}
\end{document}